\documentclass[usenatbib,usegraphicx]{mn2e}
\newcommand{\Ledd}{$L/L_{\rm Edd}$}

\newcommand{\Hb}{H$\rm\beta$}
\newcommand{\Hg}{H$\rm\gamma$}
\newcommand{\Ox}{[\mbox{O\,{\sc iii}}]}
\newcommand{\Oxx}{\mbox{O\,{\sc iii}}}
\newcommand{\Oxs}{[\mbox{O\,{\sc iii}}]~$\lambda$5007}
\newcommand{\Oxf}{[\mbox{O\,{\sc iii}}]~$\lambda$4363}

\newcommand{\Fe}{\mbox{Fe\,{\sc ii}}}
\newcommand{\He}{\mbox{He\,{\sc ii}}~$\lambda$4686}
\newcommand{\Civ}{\mbox{C\,{\sc iv}}~$\lambda$1549}

\newcommand{\rS}{$r_{\rm S}$}

\newcommand{\hst}{{\it HST}}
\newcommand{\kms}{km~s$^{-1}$}

\newcommand{\cl}{{\sc CLOUDY}}

\title[What controls the \Oxs\ line strength in AGN?]{What controls the \Oxs\
line strength in active galactic
nuclei?}
\author[A. Baskin and A. Laor]{Alexei Baskin\thanks{E-mail: alexei@physics.technion.ac.il
(AB); laor@physics.technion.ac.il (AL)} and Ari Laor\footnotemark[1] \\
Physics Department, Technion, Haifa~32000, Israel}
\begin{document}
\maketitle
\begin{abstract}
Active galactic nuclei (AGN) display an extreme range in the narrow
emission line equivalent widths. Specifically, in the PG quasar
sample the equivalent width of the narrow \Oxs\ line has a range
of $>300$ ($<0.5$\AA\ to 157\AA), while the broad \Hb\ line, for example,
has a range of ten only (23\AA\ to 230\AA). The strength of \Oxs\ is
modulated by the covering factor (CF) of the narrow line region
(NLR) gas, its density ($n_e$), and ionization parameter ($U$). To
explore which of these factors produces the observed large range
in \Oxs\ strength, we measure the strength of the matching narrow
\Hb\ and \Oxf\ lines, detected in 40 out of the 87 $z<0.5$ PG
quasars in the Boroson \& Green sample. The photoionization code
\cl\ is then used to infer CF, $n_e$, and
$U$ in each object, assuming a single uniform emitting zone. We
find that the range of CF ($\sim 0.02-0.2$)
contributes about twice as much as the range in both $n_e$ and $U$
towards modulating the strength of the \Oxs\ line. The CF is
inversely correlated with luminosity, but it is not correlated
with \Ledd\ as previously speculated. The single zone \Oxs\ emitting region
is rather compact, having $R_{\rm NLR}=40L_{44}^{0.45}$~pc.
These emission lines can also be fit with an extreme two zone
model, where \Oxf\ is mostly emitted by a dense ($n_e=10^7$~cm$^{-3}$)
inner zone at $R_{\rm NLR}^{\rm in}=L_{44}^{0.5}$~pc, and \Oxs\
by a low density ($n_e=10^3$~cm$^{-3}$) extended outer zone at
$R_{\rm NLR}^{\rm out}=750L_{44}^{0.34}$~pc. Such an extended \Oxs\
emission should be well resolved by Hubble Space Telescope imaging of luminous
AGN. Further constraints on the
radial gas distribution in the NLR can be obtained
from the spectral shape of the IR continuum emitted by the associated dust.

\end{abstract}

\section{Introduction}
Active galactic nuclei (AGN) can show both broad (FWHM$\ga
1000$~\kms) and narrow (FWHM$\la 1000$~\kms) emission lines. While
both lines are excited by the central ionizing continuum source,
the broad lines arise from the immediate vicinity of the central
massive black hole ($R \la 0.1$~pc), and the narrow lines arise
further out, sampling the host galaxy dynamics, and possibly
chemical composition and gas distribution as well. The strength
and profiles of the narrow emission lines were studied extensively
over the past 30 years, revealing radial gradients in the density
and ionization state of the gas, and some characteristic trends in
the line profiles (e.g. Heckman et al. 1981; Osterbrock 1989;
Peterson 1997; Vanden Berk et al. 2001).

One remarkable property of the narrow emission lines, which has
not been explored yet in detail, is their extreme range of
equivalent widths (EWs). Specifically, in the Boroson \&
Green (1992, hereafter BG92) sample of 87 $z<0.5$ quasars from the
bright quasar survey (Schmidt \& Green 1983), the narrow \Oxs\
line ranges in EW from 157~\AA\ to undetectable ($<0.5$~\AA), i.e.
by a factor of $>300$, while in contrast the EW of the broad \Hb\
line in the same sample ranges only over a factor of 10 (23 --
230~\AA). The narrow line EW is mainly set by two factors, the covering
factor (CF) of the photoionized gas at the narrow line region
(NLR), and the line emissivity (assuming no foreground absorption).
The CF is set by the spatial
distribution of the gas at the NLR, and by the angular
distribution of the illuminating ionizing radiation. The latter may be set by
the continuum emission mechanism, or through obscuration further
out (e.g. by the `torus'). The line emissivity for collisionaly
excited lines, such as \Oxs, is set by the gas electron density
$n_e$, the gas temperature, and by the gas ionization state. The
last two properties are set by the ionizing photon flux, or
equivalently the ionization parameter ($U\equiv n_{\gamma}/n_e$,
where $n_{\gamma}$ is the density of H ionizing photons). {\em
Which of the above factors produces the large range in EWs
of the narrow emission lines?}

The purpose of this paper is to answer this question for the \Oxs\ line.
We have chosen this line since it is generally the strongest and least
blended narrow emission line in AGN. Our approach is to measure the
strength of the narrow \Oxf\ and
\Hb\ lines in a large sample of AGN, where the \Oxs\ line was already measured.
The \Oxf/\Oxs\ line ratio is mainly a measure of $n_e$, while
the \Hb\ line is dominated by recombination and thus
mostly sensitive to the CF of the gas. The combination of the three
lines allows us to determine the CF, $n_e$, and $U$ of
the \Oxs\ emitting gas, and thus find which factor most strongly
modulates the strength of the line. This derivation relies on the
simplifying assumption that \Hb, \Oxs, and \Oxf\ originate in a single zone
with homogeneous gas properties. We relax this assumption by also making
an extreme two zone approximation, where \Oxs\ is mostly emitted by an outer
low density NLR component, and \Oxf\ by an inner high density component,
and again explore whether the strength of the \Oxs\ line is mostly
modulated by the CF or by the gas emissivity.

The \Oxs\ line is one of the main components contributing to the
set of strongly correlated AGN emission properties, so called
`eigenvector 1' (hereafter EV1, BG92). We therefore also briefly
explore whether the physical parameters controlling the strength
of \Oxs\ are linked to other EV1 emission properties. The
determination of $U$ and $n_e$ allows us to calculate the distance
of the NLR from the central ionizing continuum source, $R_{\rm
NLR}$, and we also look for relations between $R_{\rm NLR}$ and
other AGN properties. The paper is organized as follows, in
Section 2 we provide a brief theoretical justification for the set
of lines used in this analysis, and explain how the NLR parameters
are derived. In Section 3 we describe the sample used and the
narrow line measurement procedure. The results are described and
discussed in Section 4, and the main conclusions are provided in
Section 5.

\section{Theoretical approach}
The \Oxs\ and \Oxf\ lines are particularly useful for constraining
$T$ and $n_e$ of the NLR gas which produces the [\Oxx] line
emission. Both lines arise from the same element, eliminating
abundance effects. Furthermore, they both arise from the same
ionization state, eliminating ionization level corrections. Both
lines arise through collisional excitation, and although they
correspond to similar transition energy, they differ in two
respects. First, \Oxs\ arises from excitation to the $^1D_2$ level,
located 2.51~eV above the ground ($^3P_0$) level, while \Oxf\
arises from excitation to the $^1S_0$ level,  5.35~eV above the
ground level. Thus, their ratio is a useful temperature diagnostic
when $kT\la 10$~eV. In addition, the two transitions have
different radiative decay rates, leading to critical densities for
collisional dexcitation of $7.0\times10^5$~cm$^{-3}$ for \Oxs\ and
$3.3\times10^7$~cm$^{-3}$ for \Oxf. Thus, the \Oxf/\Oxs\ line
ratio is also a useful density diagnostic, when the density is not
much above or below the two critical densities (e.g. Osterbrock
1989). An additional advantage of the \Oxs\ line is that it is
typically the main coolant of the NLR (amounting to as much as
$\sim 50$ per cent of the total cooling), and thus its absolute
intensity is only a weak function of the O abundance (through the
metalicity dependence of the gas temperature).

Figure 1 presents curves of constant log~\Oxf/\Oxs\ photon flux
ratio as a function of the electron density and temperature. The
calculations are based on a 6-levels \Ox\ model (excitations to
levels above the 6th level, $^1S_0$, do not contribute significantly to the
two \Ox\ lines at the relevant temperatures). The level
energies were obtained from Moore (1993, p.~267), and radiative
transition probabilities and mean collision strengths from Pradhan
\& Peng (1995, table 1 there, for $T=10^4$~K, relevant here)
\footnote{Two collision strengths,
$\Omega$(2$p^{3~5}$S$^0_2$,~2$p^{2~1}$S$_0$) and
$\Omega$(2$p^{3~5}$S$^0_2$,~2$p^{2~1}$D$_2$), were not available,
and we therefore assume are equal to 1.}. The figure demonstrates
the weaker dependence of the line ratio on $n_e$ at the
lowest $n_e$, and weaker dependence on $T$ as the highest $T$.

In photoionized gas it is more natural to use $U$, which is
often a free parameter in photoionization calculations, rather than $T$
which is set by $U$ and $n_e$ through detailed photoionization
calculations. Fig.~1 presents the dependence of $T$ on $n_e$ and
$U$, as deduced from the photoionization code \cl\ (Ferland et al.
1998). It demonstrates the weak dependence of $T$ on $n_e$, at a
given $U$, and the overall small range in $T$ ($1$-$2.5\times
10^4$~K) when $U$ varies from $10^{-4}$ to $10^{-1}$. Assuming a
uniform $n_e$ gas, the measured value of \Oxf/\Oxs\ (corresponding to a
certain curve in Fig~.1) yields the allowed range of $n_e$ as a
function of the range in $U$. E.g., if the measured line ratio is
0.1, then the implied density is
$n_e=10^{5.75}$ to $10^{6.3}$~cm$^{-3}$, for $U=10^{-1}$ to $10^{-4}$.
As can be seen in Fig.~1, the allowed range in $n_e$ increases as
the line ratio gets smaller.

Breaking the $n_e$, $U$, degeneracy requires an additional line.
The optimal line for that purpose is \Hb. It is a nearly pure
recombination line at the expected NLR conditions (where
collisional and optical depth effects should be small), and thus
its intensity provides an estimate of the fraction of ionizing
radiation intercepted by the NLR, i.e. the NLR CF. The \Oxs/\Hb\
ratio provides the additional constraint on $n_e$ and $U$,
required to break the degeneracy. The \Hb\ EW, together with the
EW of \Oxs, and \Oxf, allow us to determine $n_e$, $U$, and the CF
of the \Ox\ emitting gas in the NLR. From the observational aspect
the \Hb\ line is also optimal, as it is located close to the two
\Ox\ lines, allowing all three lines to be obtained in a single
spectrum, minimizing systematic measurement errors. Also, the
close wavelength proximity of the lines minimizes dust reddening
effects on the measured line ratios.

As noted above, we use the photoionization code \cl\ to calculate
the strength of \Oxs, \Oxf, and \Hb, as a function of $n_e$
(from $10^3$~cm$^{-3}$ to $10^8$~cm$^{-3}$), $U$ (from $10^{-4}$ to
$10^{-1}$),
and the CF. The additional input model parameters for \cl\ are a
slab geometry with a total column of $10^{23}$~cm$^{-2}$, ISM
abundances with grains, and a standard Mathews \& Ferland (1987;
herafter MF)
AGN continuum with a break at 1~$\mu m$ (Ferland et al. 1998).
The results are not sensitive to the assumed column since the dust
opacity strongly suppresses line emission from columns beyond
$10^{21}$~cm$^{-2}$ (Laor \& Draine 1993, hereafter LD).
The effect of the ionizing continuum shape
is explored by calculating the model results again with a different
ionizing spectrum, as further described in Section~4.

Figure 2 presents the
calculated line flux ratios and EW as a function of $n_e$ for
different values of $U$. The upper panel shows the \Oxs/\Hb\
flux ratio. This ratio decreases with increasing $n_e$ due to
collisional suppression of the \Oxs\ line at $n_e>10^{5.85}$.
The ratio increases with $U$ at
low $U$, as the fractional abundance of \Oxx\ increases, and
starts decreasing at $\log U>-1.5$, as O gets ionized beyond \Oxx\
(the dust opacity suppresses emission from deeper layers where
O is less ionized).
The middle panel shows the \Oxf/\Oxs\ flux ratio. This ratio is mostly
set by $n_e$ due to the different $n_{\rm crit}$ of the two
lines, as discussed above. The weak dependence on $U$ results from
the weak dependence of $T$ on $U$ (Fig.~1). The lower panel shows
the \Hb\ EW for a CF~=~1. At low $U$ the \Hb\ EW is independent of
$n_e$ and $U$, as expected for a pure recombination line. At $\log
U>-2.5$ the \Hb\ EW starts decreasing due to the increasing
suppression of the line emission by absorption in dust within the
ionized gas (Voit 1992; LD; Netzer \& Laor 1993; Ferguson et al. 1997;
Dopita et al. 2002).

Figure 3 presents theoretical curves of the \Oxs/\Hb\ flux ratio as a
function of the \Oxf/\Oxs\ flux ratio for different values of $n_e$ and
$U$. This plot allows one to transform the two measured line
ratios in a given object to the corresponding values of $n_e$ and
$U$ in that object. The third parameter, the CF, is then set by
the EW of either one of the three lines. There is a one to one
transformation from the \Oxs/\Hb\ versus \Oxf/\Oxs\ coordinate to
the $n_e$ versus $U$ coordinate, as long as $\log U\la -2.2$.
Above that value there are two sets of solutions, one having $\log
U< -1.5$, and the other $\log U> -1.5$, for a given set of
\Oxs/\Hb\ and \Oxf/\Oxs\ values (see Fig.~3). The presence of two
solutions results from the
non-monotonic dependence of \Oxs/\Hb\ on $U$, mentioned above
(Fig.~2, upper panel). The two solutions correspond to different CF values
(the larger $U$ requires a larger CF to produce the observed \Hb\ EW,
Fig.~2 lower panel). Obtaining a unique solution requires
additional constraints on $U$ or CF. As further described in
Section 3, almost none of our objects happens to lie within the
degenerate solution regime. However, note that photoionization models
with $\log U>-1$, not explored here, produce lower \Oxs/\Hb\ ratios,
which will be degenerate with some $\log U\la -2.2$ solutions. Such high
$U$ models require a high CF, and they imply strong high ionization
narrow lines, which we do not study here.

The above analysis is within the single zone approximation, i.e.
all three lines are assumed to originate in gas with a given set
of $n_e$, $U$ and CF. In reality, there is likely to exist a distribution of
values for these three parameters in the NLR, in which case the
relative line strength will be a function of position in the NLR (e.g. Ferguson
et al. 1997).
The assumption of a single zone with uniform gas properties
may not yield representative mean values in this case. To explore the effects of
a non uniform cloud population, we repeated the analysis assuming
two populations of clouds with very different \Oxs\ and \Oxf\
emissivities. The first population is assumed to have a density
 $n_e=10^7$~cm$^{-3}$, and thus produces a high \Oxf/\Oxs\ ratio,
and the second population is assumed to have $n_e=10^3$~cm$^{-3}$,
and thus produces a ratio lower by a factor of $\sim 100$
(Fig.~3). Most of the \Oxf\ emission is expected
to come from the first population of clouds, and most \Oxs\ from the second.
Such a ``two zone model"  has six free parameters ($n_e$, $U$, and CF for
each zone), and three need to be set since we only have three
constraints (the EWs of \Hb, \Oxs, and \Oxf). The additional parameter
we fix is $\log U=-1$ for the $n_e=10^7$~cm$^{-3}$ zone, which implies this
is an inner very compact zone. This rather high $U$ is motivated by the mean
spectral energy distribution (SED) of AGN, which indicates a
significant covering factor of very hot
dusty gas with a high ionization parameter. The dust temperature indicates
it is located outside the BLR, but
on scales smaller than the ``typical" NLR. The lack of strong line
emission from the associated gas implies $U\gg 10^{-2}$, to allow
significant suppression of the line emission by the dust (Section 2).
The remaining free parameters
of the two zone model are the ionization parameter of the outer zone, $U_{\rm out}$,
and the covering factors of the inner and outer zones,
CF$_{\rm in}$, CF$_{\rm out}$, these are set as described below
in Section 3.4.

\begin{figure}
\includegraphics[width=84mm]{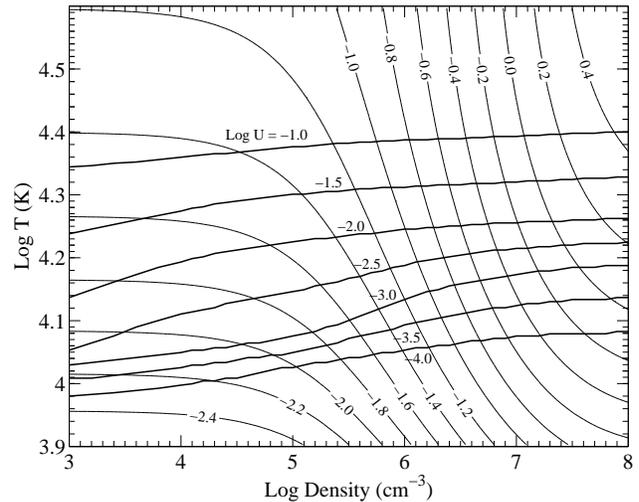}
\caption{Curves of constant log~\Oxf/\Oxs\ photon flux ratio as a
function of log $n_e$ and log $T$ for a 6-levels collisionaly
excited \Oxx. The thick lines represent $T(n_e)$ for different
values of $U$, as calculated by the photoionization code \cl. Note
the weak dependence of $T$ on $n_e$ at a given $U$.}
\end{figure}

\begin{figure}
\includegraphics[width=84mm]{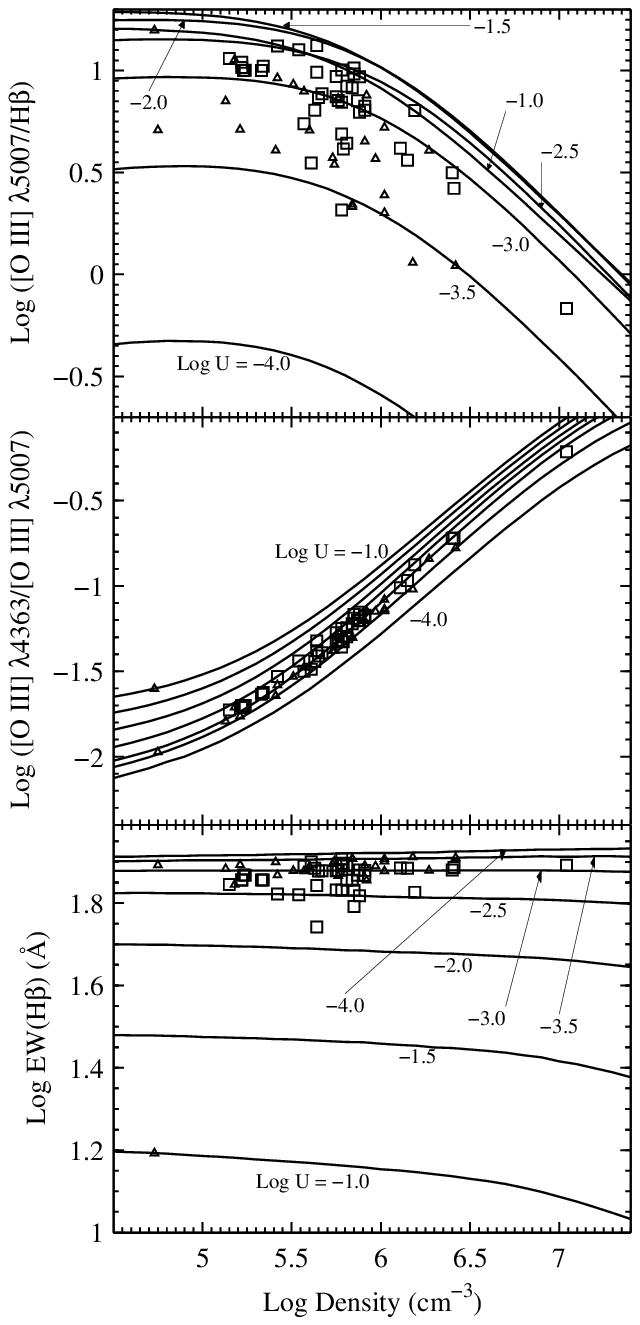}
\caption{The solid lines present the dependence of \Oxs/\Hb\
(upper panel), \Oxf/\Oxs (middle panel), and \Hb\ EW (lower panel)
on $n_e$, at given values of $U$, as calculated by \cl. The square
boxes represents objects with detections, and the triangles objects
with an upper limit on the \Oxf\ EW, i.e. an upper limit on
$n_e$. In the lower panel the calculated \Hb\ EW is for CF~=~1,
and for each object we therefore plotted the
observed \Hb\ EW divided by the inferred CF.}
\end{figure}

\begin{figure}
\includegraphics[width=84mm]{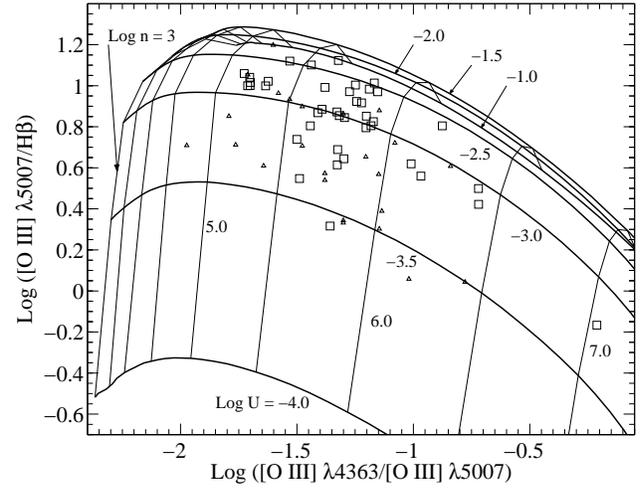}
\caption{The theoretical and measured relation between \Oxs/\Hb\
and \Oxf/\Oxs. The solid lines represent curves of constant $n_e$
and constant $U$, as calculated by \cl. The object symbols are as
in Fig.~2.}
\end{figure}

\section{Data Analysis}
\subsection{The data set}
For the analysis we use the BG92 sample mentioned
above (Section 1). This sample extends in luminosity from Seyfert
galaxies with $\nu L_{\nu}=3.3\times 10^{43}$~erg~s$^{-1}$
(calculated at rest frame 3000~\AA\ using the continuum fluxes in
Neugebauer et al. 1987, assuming $H_0=80$~km~s$^{-1}$~Mpc$^{-1}$,
$\Omega_0=1.0$), to luminous quasars at $\nu L_{\nu}=1.4\times
10^{46}$~erg~s$^{-1}$. This is a complete and well defined sample,
selected based on (blue) color and (point like) morphology,
independently of the emission line strengths and profiles
(subject to the condition that broad emission lines are
present).

Optical observations of the \Hb\ region of the 87 objects are
described in BG92, and were kindly provided  by T.
Boroson (private communication). These spectra do not
extend down to \Oxf\ in all objects, and we therefore
supplemented the BG spectra by optical spectra from Shang et~al.~(2003),
kindly provided by Z. Shang (private communication) which also provide
a higher S/N in some cases\footnote{The spectra obtained from
Shang et al. (2003) are of PG~0953+414, PG~1216+069, PG~1322+659,
PG~1427+480, PG~1512+370 and PG~1543+489.}. The final sample contains
78 objects with spectra which contain the \Oxf\ region.
The spectra were corrected for reddening, redshift, and possible slit
losses, as described in Baskin \& Laor (2005, hereafter BL05,
Section 2.2 there).

\subsection{The \Fe\ subtraction}
The \Oxf\ line is generally very weak, and its measurement requires
a careful subtraction of the \Fe\ emission multiplets. We used
the I~Zw~1 \Fe\ template, kindly provided by T. Boroson, to
subtract the \Fe\ lines from the spectra (see Section 2.2 in
BL05 for further details concerning the subtraction procedure).

The \Fe\ template was constructed by subtracting all the none
\Fe\ emission from the observed spectrum of I~Zw~1. However,
this also requires estimates of the relative contributions of \Fe\
and \Oxf\ near 4363\AA, which may not be accurate, and could lead
to biased estimates of the \Oxf\ strength in the \Fe\ subtracted
spectra. To check the accuracy of the \Fe\ subtraction, we also
used the more
theoretically motivated \Fe\ template of I Zw 1, independently
constructed by V{\' e}ron-Cetty, Joly \& V{\' e}ron (2004),
kindly provided by M.~V{\' e}ron-Cetty. The BG92 template is rather
smooth and featureless around 4363\AA, and the V{\' e}ron-Cetty et al.
template has a sharp feature redward of 4363\AA, but much weaker emission
at 4363\AA, compared to BG92.
The V{\' e}ron-Cetty et al. compilation of expected allowed and forbidden \Fe\
emission lines indicates no significant feature near 4363\AA.
There was almost no effect on the \Oxf\ profile in
I~Zw~1 when using either the V{\' e}ron-Cetty et al. template or
the BG92 template to subtract the \Fe\ emission. We therefore conclude
that improper \Fe\ subtraction is unlikely to significantly bias
our measurements of the \Oxf\ line. In our analysis we used the
empirically derived BG92 \Fe\ template.

\subsection{The measurement procedure}
As noted above, there is likely to exist a distribution
of $n_e$, $U$, and CF values for the \Ox\ emitting gas in the NLR.
To ensure as much as possible that the measured \Hb\ and \Oxf\
emission originates in the same gas which produces the \Oxs\
emission, we use the \Oxs\ profile as a template for measuring
the \Hb\ and \Oxf\ lines. This ensures that all three lines
originate in gas with the same line of sight velocity distribution,
although it does not necessarily guarantee they all originate in
NLR gas clouds with the same distribution of emission properties.

In each object we subtracted the maximum allowed scaled \Oxs\
profile which does not produce a dip in the broad \Hb\ line and in the
\Hg+\Oxf\ blend, as determined by eye inspection.
The subtraction was performed as follows. Both the \Oxs\ line and
either \Oxf\ or \Hb\ were transformed from wavelength scale
to velocity scale, using $z$
measured from the \Oxs\ peak (T. Boroson, private communication,
listed in BL05). Then, the maximal $c$
in the term $f_{\rm line}(v)-c\times f_{5007}(v)$ was searched,
where `line' is either \Hb\ or \Oxf,
such that the subtraction of the scaled \Oxs\ profile
will not produce a dip in the measured line. The measured narrow
line component equals the scaled \Oxs\ line,
$f_{\rm line}(v)=c\times f_{5007}(v)$. To calculate the
line EW the line profile $f_{\rm line}(v)$
was transformed back to wavelength dependence $f_{\rm line}(\lambda)$,
integrated over $\lambda$, and divided by
the continuum flux density at 4861~\AA\ (the same wavelength used by BG92).
The flux density at 4861~\AA\ was measured by fitting
a local power-law
continuum to each spectrum between $\sim4600$~\AA\ and
$\sim5100$~\AA. We measured the line EW directly,
rather than infer it from the \Oxs\ EW measured by BG92
and the value of $c$, since the \Oxs\ profile deduced here
may not be identical to the one assumed by BG92 due to
possibly different definitions of the continuum underlying \Oxs.

To determine whether the measured line flux is significant or not,
we need to evaluate the standard deviation, $\sigma$, of the
flux measurement.
We first measured the RMS of the flux density between 4500\AA\ and
4600\AA\footnote{Seven spectra have a feature
in this range, and thus a slightly different range was chosen for
them.}, multiplied the RMS by the wavelength width of a pixel (to convert
flux density to flux), and then multiplied it
by the square-root of the number of pixels in the full width at zero
intensity of the \Oxs\ profile, to get $\sigma$.
A measured line flux was accepted as significant if it is $>3\sigma$.
Following BG92, we also adopted an EW of 0.5~\AA\ as the lower detection
limit (accounting for the minimum likely level of systematic errors).
After exclusion of objects with an unmeasurable \Hb\ narrow component,
and an unmeasurable \Oxf, the remaining sample
contained 40 objects where all three lines are detected.

Figure 4 shows four representative cases of the \Oxs\
profile subtraction from the \Hg+\Oxf\ blend. The two
left panels present objects where the \Ox~$\lambda$4363 emission
is rather well fit using the \Oxs\ profile.
The upper left panel is a clear $18.6\sigma$ detection, and the lower
left one a marginally acceptable detection at $3.6\sigma$. The two
right panels in Fig.~2 demonstrate cases where a broad residual
\Oxf\ component remains after subtraction of the
scaled \Oxs\ profile. Such a broad residual component
often remains, and it most likely indicates an inner higher density
extension of the NLR, where the density is higher than the \Oxs\ critical
density but it is lower than the critical density of \Oxf. Such an inner,
higher velocity, extension of the NLR would contribute to
the \Oxf\ line and not to the \Oxs\ line (Nagao et al. 2001).
This demonstrates that
making inferences based on the total \Oxs\ and \Oxf\ line fluxes,
rather than matched profile fluxes, could lead to inaccurate results.

\begin{figure}
\includegraphics[width=84mm]{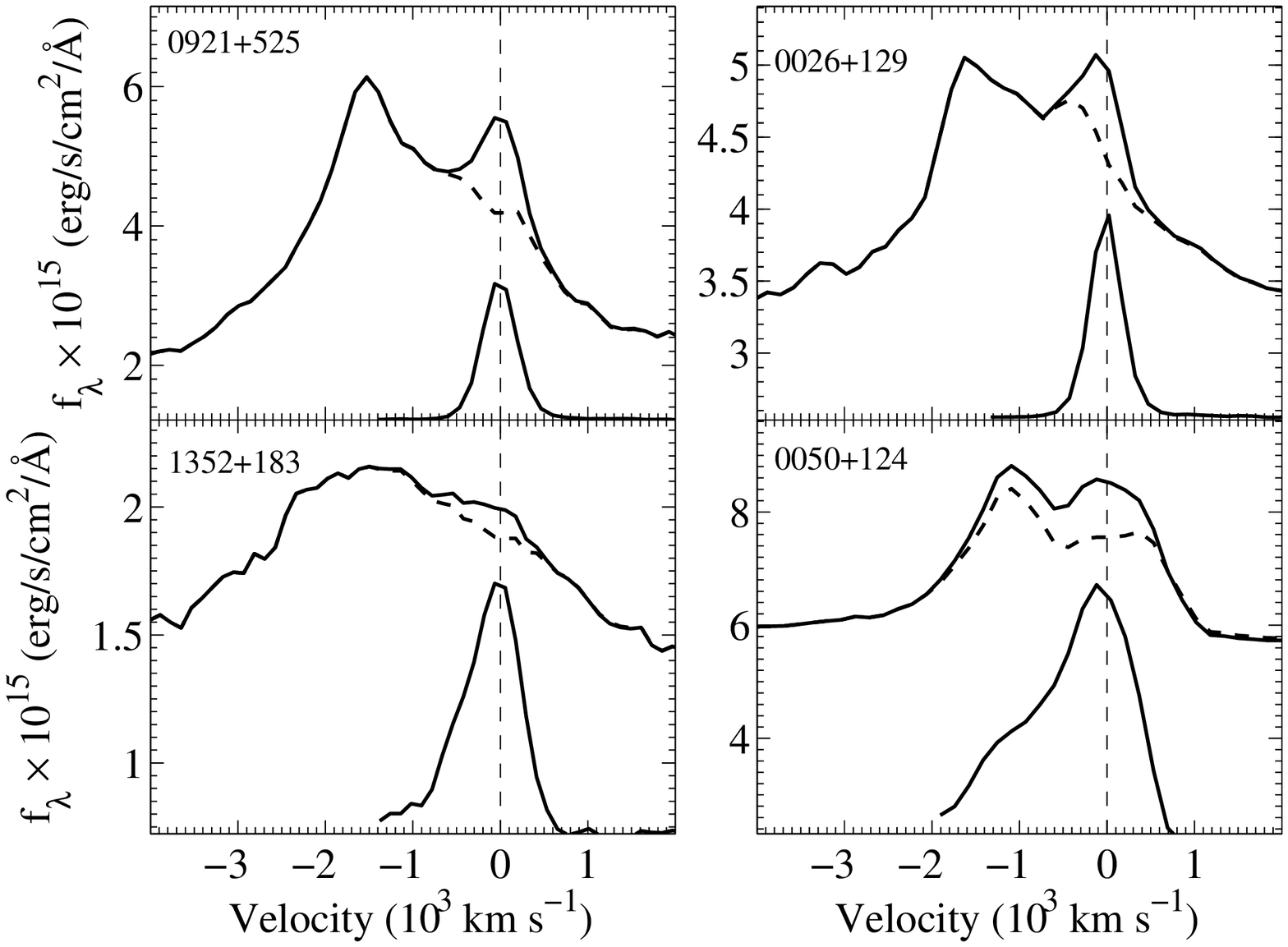}
\caption{The \Hg\ and \Oxf\ blend before (top
continuous line) and after (dashed line) substraction of the
scaled \Oxs\ profile, and the \Oxs\
profile (scaled by an arbitrary factor for presentation; bottom
continuous line). The name of the object is indicated in each panel.
Upper-left, the significance of the measurement is $18.6\sigma$.
Lower-left, the significance is $3.6\sigma$. The two right
panels demonstrate a broad residual \Oxf\ component
that often remains (see text).}
\end{figure}

\subsection{The inferred NLR parameters}

Table 1 presents the narrow ``\Oxs-like"
\Hb\ and \Ox~$\lambda$4363 EW, measured as
described above.  We also list in Table 1 the \Oxs\ EW from BG92
(the Shang et~al. spectra yielded consistent values).
Column 1 lists the name of the object,
columns 2 and 3 list the measured \Hb\ EW in units of \AA\ and its
significance in units of $\sigma$. Column 4 lists the \Oxs\ EW in
units of \AA, as reported by BG92. Columns 5 and 6 list the
measured \Oxf\ EW and its significance. Column 7 lists $\log\nu
L_{\nu}$ at 4861~\AA\ in units of $10^{44}$~erg~s$^{-1}$, which
allows a quick conversion of line EW to line luminosities. Columns
8, 9 and 10 list the inferred NLR parameters, $\log U, \log n_e$
(in units of cm$^{-3}$), and CF (in per cent) for the single-zone
model. Finally, columns 11, 12 and 13 list the CF$_{\rm in}$ (in
per cent), $\log U_{\rm out}$ and CF$_{\rm out}$ (in per cent) for
the two-zones model.

\begin{table*}
\begin{minipage}{142mm}
\caption{The measured emission line parameters and the inferred
NLR parameters.} 
\begin{tabular}{@{}l*{12}{r}@{}}
\hline Object & \multicolumn{2}{c}{\Hb} &
\multicolumn{1}{c}{\Ox}&\multicolumn{2}{c}{\Ox~$\lambda$4363} &
\multicolumn{1}{c}{$\nu L_{\nu}$}& \multicolumn{3}{c}{Single zone}
& \multicolumn{3}{c}{Two zones}
 \\
& \multicolumn{1}{c}{EW} & \multicolumn{1}{c}{S/N} &
\multicolumn{1}{c}{$\lambda$5007} &
\multicolumn{1}{c}{EW} & \multicolumn{1}{c}{S/N} & & \multicolumn{1}{c}{$U$} & \multicolumn{1}{c}{$n$}
& \multicolumn{1}{c}{CF} & \multicolumn{1}{c}{CF$_{\rm in}$} & \multicolumn{1}{c}{$U_{\rm out}$} & \multicolumn{1}{c}{CF$_{\rm out}$}\\
\multicolumn{1}{c}{(1)} & \multicolumn{1}{c}{(2)} &
\multicolumn{1}{c}{(3)} & \multicolumn{1}{c}{(4)} &
\multicolumn{1}{c}{(5)} & \multicolumn{1}{c}{(6)} &
\multicolumn{1}{c}{(7)} & \multicolumn{1}{c}{(8)} &
\multicolumn{1}{c}{(9)} & \multicolumn{1}{c}{(10)}&
\multicolumn{1}{c}{(11)}& \multicolumn{1}{c}{(12)}&
\multicolumn{1}{c}{(13)}\\ \hline
0003+158    &   1.96    &   20.9   &   26  &   1.24    &   16.5   &   $   1.579   $   &   $   -2.2    $   &   5.64    &   3.6 &   4.8 &   $   -1.50   $   &   4.7 \\
0003+199    &   5.34    &   22.9   &   22  &   1.04    &   6.8    &   $   -0.267  $   &   $   -3.3    $   &   5.79    &   6.8 &   4.8 &   $   -3.18   $   &   6.1 \\
0007+106    &   4.19    &   16   &   42  &   0.84    &   4.4    &   $   0.407   $   &   $   -2.9    $   &   5.24    &   5.7 &   2.8 &   $   -2.45   $   &   5.8 \\
0026+129    &   3.10    &   34.9   &   29  &   1.54    &   21.2   &   $   0.804   $   &   $   -2.6    $   &   5.75    &   4.6 &   6.9 &   $   -2.23   $   &   3.9 \\
0049+171    &   15.55   &   18.8   &   97  &   6.07    &   10.9   &   $   -0.509  $   &   $   -3.0    $   &   5.88    &   20.5    &   28.9    &   $   -2.85   $   &   16.4    \\
0050+124    &   6.97    &   41.4   &   22  &   4.19    &   27.5   &   $   0.312   $   &   $   -3.0    $   &   6.40    &   9.2 &   21.4    &   $   -3.22   $   &   5.6 \\
0052+251    &   3.90    &   29.9   &   30  &   1.22    &   10.3   &   $   0.808   $   &   $   -3.0    $   &   5.67    &   5.2 &   5.4 &   $   -2.68   $   &   4.6 \\
0838+770    &   3.58    &   8.8    &   13  &   1.40    &   4.4    &   $   0.356   $   &   $   -3.1    $   &   6.15    &   4.7 &   7.0 &   $   -3.20   $   &   3.5 \\
0921+525    &   13.00   &   33.3   &   83  &   5.46    &   18.6   &   $   -0.659  $   &   $   -3.0    $   &   5.91    &   17.2    &   26.0    &   $   -2.80   $   &   13.5    \\
0923+129    &   7.04    &   24.7   &   31  &   1.56    &   8.2    &   $   -0.579  $   &   $   -3.2    $   &   5.81    &   9.1 &   7.3 &   $   -3.15   $   &   8.1 \\
0934+013    &   7.29    &   22.6   &   51  &   2.58    &   11.7   &   $   -0.603  $   &   $   -3.0    $   &   5.78    &   9.6 &   11.9    &   $   -2.75   $   &   8.2 \\
0953+414    &   1.87    &   10.0   &   18  &   1.16    &   7.2    &   $   1.033   $   &   $   -2.6    $   &   5.85    &   2.8 &   5.3 &   $   -2.01   $   &   2.3 \\
1001+054    &   2.65    &   5.4    &   7   &   1.33    &   3.2    &   $   0.480   $   &   $   -3.1    $   &   6.41    &   3.4 &   6.8 &   $   -3.34   $   &   2.3 \\
1048+342    &   2.72    &   5.9    &   28  &   1.90    &   5.5    &   $   1.235   $   &   $   -2.4    $   &   5.85    &   4.4 &   8.4 &   $   -1.71   $   &   4.4 \\
1100+772    &   3.90    &   15.8   &   41  &   0.98    &   4.4    &   $   1.276   $   &   $   -2.8    $   &   5.34    &   5.4 &   3.5 &   $   -2.32   $   &   5.6 \\
1119+120    &   2.66    &   23.0   &   19  &   0.91    &   9.7    &   $   -0.140  $   &   $   -3.0    $   &   5.76    &   3.5 &   4.2 &   $   -2.73   $   &   3.0 \\
1126$-$041  &   3.47    &   12.4   &   19  &   0.60    &   3.2    &   $   0.021   $   &   $   -3.2    $   &   5.57    &   4.5 &   2.6 &   $   -3.05   $   &   4.1 \\
1149$-$110  &   4.64    &   16.7   &   33  &   2.07    &   9.5    &   $   -0.392  $   &   $   -2.9    $   &   5.87    &   6.3 &   9.8 &   $   -2.66   $   &   4.9 \\
1151+117    &   1.33    &   17.1   &   11  &   0.66    &   10.6   &   $   0.456   $   &   $   -2.8    $   &   5.84    &   1.9 &   3.1 &   $   -2.44   $   &   1.5 \\
1202+281    &   5.64    &   37.3   &   36  &   1.30    &   9.8    &   $   0.386   $   &   $   -3.1    $   &   5.63    &   7.4 &   5.7 &   $   -2.93   $   &   6.6 \\
1229+204    &   1.94    &   10.9   &   19  &   0.79    &   6.5    &   $   0.067   $   &   $   -2.7    $   &   5.64    &   2.8 &   3.4 &   $   -2.26   $   &   2.6 \\
1244+026    &   8.21    &   28.1   &   17  &   0.75    &   3.5    &   $   -0.454  $   &   $   -3.5    $   &   5.78    &   10.2    &   3.5 &   $   -3.51   $   &   9.6 \\
1302$-$102  &   2.16    &   12.8   &   9   &   0.88    &   5.5    &   $   1.510   $   &   $   -3.1    $   &   6.11    &   2.8 &   4.4 &   $   -3.13   $   &   2.1 \\
1307+085    &   3.18    &   14.7   &   32  &   1.80    &   11.8   &   $   0.716   $   &   $   -2.6    $   &   5.78    &   4.7 &   8.0 &   $   -2.01   $   &   4.2 \\
1309+355    &   2.55    &   18.0   &   19  &   0.89    &   7.1    &   $   0.756   $   &   $   -3.0    $   &   5.75    &   3.4 &   4.1 &   $   -2.68   $   &   2.9 \\
1310$-$108  &   8.24    &   39.2   &   77  &   5.41    &   30.6   &   $   -0.637  $   &   $   -2.5    $   &   5.88    &   12.5    &   24.8    &   $   -2.01   $   &   10.0    \\
1341+258    &   1.90    &   9.8    &   14  &   0.54    &   4.1    &   $   -0.053  $   &   $   -3.0    $   &   5.65    &   2.5 &   2.4 &   $   -2.73   $   &   2.2 \\
1351+236    &   2.46    &   11.2   &   12  &   0.57    &   3.0    &   $   -0.204  $   &   $   -3.2    $   &   5.78    &   3.2 &   2.6 &   $   -3.09   $   &   2.8 \\
1351+640    &   3.70    &   20.6   &   31  &   1.77    &   11.3   &   $   0.515   $   &   $   -2.8    $   &   5.81    &   5.2 &   8.2 &   $   -2.44   $   &   4.1 \\
1352+183    &   1.57    &   4.8    &   10  &   1.33    &   3.6    &   $   0.432   $   &   $   -2.6    $   &   6.19    &   2.3 &   6.6 &   $   -2.34   $   &   1.2 \\
1411+442    &   2.25    &   9.9    &   15  &   1.01    &   6.5    &   $   0.279   $   &   $   -2.9    $   &   5.91    &   3.1 &   4.8 &   $   -2.73   $   &   2.3 \\
1425+267    &   3.61    &   20.0   &   36  &   0.84    &   4.9    &   $   1.301   $   &   $   -2.8    $   &   5.33    &   5.0 &   3.0 &   $   -2.42   $   &   5.0 \\
1501+106    &   5.58    &   35.6   &   64  &   1.20    &   10.9   &   $   0.016   $   &   $   -2.7    $   &   5.15    &   8.0 &   3.6 &   $   -2.22   $   &   8.8 \\
1512+370    &   5.50    &   20.9   &   57  &   1.14    &   5.2    &   $   0.958   $   &   $   -2.8    $   &   5.22    &   7.7 &   3.8 &   $   -2.39   $   &   7.9 \\
1519+226    &   5.87    &   8.6    &   4   &   2.45    &   4.8    &   $   0.265   $   &   $   -3.2    $   &   7.04    &   7.5 &   12.8    &   $   -3.96   $   &   5.4 \\
1534+580    &   5.99    &   48.3   &   79  &   2.33    &   25.3   &   $   -0.542  $   &   $   -2.5    $   &   5.42    &   9.0 &   7.8 &   $   -1.75   $   &   12.5    \\
1535+547    &   4.53    &   22.0   &   16  &   0.52    &   4.3    &   $   -0.286  $   &   $   -3.4    $   &   5.61    &   5.7 &   2.3 &   $   -3.27   $   &   5.4 \\
1545+210    &   2.61    &   11.2   &   33  &   1.20    &   6.5    &   $   1.093   $   &   $   -2.5    $   &   5.54    &   4.0 &   4.4 &   $   -1.72   $   &   5.3 \\
1612+261    &   14.28   &   74.3   &   157 &   3.11    &   23.1   &   $   0.332   $   &   $   -2.8    $   &   5.22    &   19.9    &   9.9 &   $   -2.28   $   &   21.6    \\
1704+608    &   2.70    &   15.2   &   27  &   0.52    &   3.4    &   $   1.410   $   &   $   -2.9    $   &   5.23    &   3.7 &   1.8 &   $   -2.46   $   &   3.7 \\
\hline
\end{tabular}
\end{minipage}
\end{table*}

To infer $U, n_e$ and CF we calculated theoretical values of the
\Hb\ flux, and \Oxs/\Hb\ and \Oxf/\Oxs\ ratios using the
photoionization code \cl\ (v.~94.00). The calculations were made
for a grid of models with $-4\le \log U\le -1$ (in steps of 0.5),
and $3\le \log n_e \le 8$ (in steps of 0.1). The \Hb\ flux was
converted to EW as follows. The choice of $U$ and $n_e$
corresponds to a given ionizing flux at the face of the
photoionized slab, which is specified by the output of \cl. We
multiplied the H ionizing flux by 0.4 to get an approximate value
for $\nu f_{\nu}$ at $\lambda 4861$ (deduced from the spectral
shape in fig.~7 of LD), and divided it by 4861 to
get the continuum $f_{\lambda}$ at 4861~\AA. The \Hb\ line flux
from \cl\ was then divided by this $f_{\lambda}$ to get the
predicted line EW for CF~=~1. This procedure assumes a uniform
spectral energy distribution (SED) in all objects, which may not
correspond well with the range of observed SEDs in the PG sample
quasars. However, as shown in Section 4.1.3, the likely range of
SED shapes has only a minor effect on the deduced parameters.

For the single zone model we use the following algorithm
to infer $U, n_e$ and CF from the
\Hb\ EW, \Oxs/\Hb\ and \Oxf/\Oxs\ ratios.
\begin{enumerate}
\item Assume $U$, starting at $\log U=-4$.
\item Calculate $n_e$ from the observed \Oxf/\Oxs\ ratio
for the assumed $U$ (Fig.~2, middle panel). The calculation
is made using a linear interpolation scheme for
log~\Oxf/\Oxs\ versus log~$n_e$, when the observed \Oxf/\Oxs\
value is between grid points.
\item Calculate the theoretical \Oxs/\Hb\ ratio for the assumed $U$,
and the $n_e$ calculated above (using a linear interpolation
scheme as above).
\item Calculate the theoretical \Oxs\ EW from the theoretical
\Oxs/\Hb\ ratio found above, and the observed \Hb\ EW.
\item Repeat steps (i)-(iv) changing $\log U$ from $-4$ to $-1$
in steps of 0.1 (using linear interpolations between $U$
grid points).
\item Adopt as the solution the values of $U$ and $n_e$ which
yield the smallest
difference between the observed and theoretical \Oxs\ EW,
as calculated in step (iv).
\item Calculate the theoretical \Hb\ EW for the $U$
and $n_e$ of the solution (Fig.~2, lower panel).
Obtain CF by dividing the observed \Hb\ EW by the
theoretical one.
\end{enumerate}
This procedure is effectively identical to the one described in
Section 2. The differences between the observed and calculated
\Oxs\ EW are typically well below 10 per cent, which is generally
within the \Oxs\ EW measurement errors. The
resulting best fit $U, n_e$ and CF for each object are listed
in Table 1.

As discussed in Section 2, in the two zone model we assume
$n_e=10^3$~cm$^{-3}$ for the
outer zone, and $n_e=10^7$~cm$^{-3}$ and $U=0.1$ for
the inner zone, leaving CF$_{\rm in}$, CF$_{\rm out}$ and $\log
U_{\rm out}$ as the remaining three parameters to be set by the
observations. The assumed $n_e$ and $U$ for the inner zone imply
log~\Oxs/\Hb=0.2, log~\Oxf/\Oxs=$-$0.04 (see Fig.~3) and a calculated
\Hb~EW of 12.21~\AA\ for CF$_{\rm in}$=1. We now iterate over the possible
range of log~$U$ values for the outer zone. For each $U$ value for
the outer zone we calculate \Oxs/\Hb\, \Oxf/\Oxs, and the \Hb\ EW for
CF$_{\rm out}$=1. We now write two linear equations
for the observed \Oxs\ and \Oxf\ EWs in a given object, as a function
of the above line ratios, the calculated \Hb~EWs, and the two
unknowns, CF$_{\rm in}$ and CF$_{\rm out}$, which we solve for.
The solution produces the correct observed \Oxs/\Hb\, \Oxf/\Oxs\
ratios by construction, and it also allows us to calculate the
total expected \Hb\ EW. The accuracy of the solution is measured
by the difference between the
observed and calculated total \Hb\ EW. The $U$ value which minimizes
this difference (typically less than 10 per cent) is taken as the
best fit solution for the given object.

\section{Results \& Discussion}

\subsection{The distribution of $n_e$ and $U$ values}
\subsubsection{The Single Zone Approximation}
As shown in Table 1 and Fig.~2, the values of $n_e$ of all 40
objects with detections (excluding one outlier, PG~1519+226) lie within
a relatively small range (factor of $\sim 5$) around the critical density
of \Oxs\ (i.e. within $\log n_e=5.85\pm 0.7$). This is expected if there
is a range of
densities in the NLR, since the line emissivity per unit mass
peaks close to the critical density, and thus the \Oxs\ emissivity
weighted mean density is expected to be at $n_e\sim n_{\rm crit}$,
unless the distribution of CF with density is heavily weighted
towards very high or very low densities.

We also find that all objects (except one) lie at $\log U\le -2.2$.
A similar
selection effect may be present here as well. As Fig.~2 (lower
panel) shows, the \Hb\ line emissivity per unit solid angle
(=~EW/CF) drops at $\log U> -2.5$ (due to dust suppression), and
thus if there is a distribution of $U$ values in the NLR with comparable
values of CF, most of
the \Hb\ emission would originate in gas with $\log U< -2$.
However, at $\log U< -3.5$ the \Oxx\ emission is suppressed
because Oxygen becomes less ionized than \Oxx. Thus, the combined
effects of dust suppression and ion abundance leads to a
relatively small range of $U$ ($\log U=-3\pm 0.5$) where both the
\Ox\ lines and \Hb\ are emitted efficiently.

In the above analysis we considered only objects where all the
three narrow lines are detected. Ignoring upper limits may result
in spurious trends or correlations. To explore that, we repeated
the analysis described in Section 3 for all the objects where
\Oxf\ was not detected, assuming the line EW equals the upper
limit value (the larger of the 3$\sigma$ detection limit and
0.5~\AA). These objects are also plotted in Figs.~2 and 3, and
their distribution nearly overlaps the distributions of the
detected objects. We also repeated the analysis arbitrarily assuming the
\Oxf\ EW is a factor of 10 lower than the upper limit. This
leads to a tail of objects extending to $\log
n_e=3$, but the distribution of $U$ values is not much affected.
The extension to low $n_e$ occurs because $n_e$ is largely set by
\Oxf/\Oxs, while $U$ is unchanged because it is largely set by
\Oxs/\Hb, and it is nearly independent of the \Oxf\ EW when
\Oxf/\Hb\ is low (Fig.~3).

\subsubsection{The Two Zone Approximation}
As described above, the two zone approximation we use assumes fixed
values for $n_e$ in the two zones, and a fixed $U$ for the inner
zone. For $U_{\rm out}$ we find $-4\le \log U_{\rm out}\le -1.5$, with most
objects in the range of $-3.5$ to $-2$. This
distribution is offset to somewhat higher $U$ values (by $\sim 0.2$)
compared to the single zone approximation (Fig.~5).

\subsubsection{Modified SED}

Far UV observations of high $z$ quasars (Zheng et al. 1997), together
with soft X-ray observations of low $z$ quasars (Laor et al. 1997),
suggest a significantly softer ionizing SED than the one suggested
by MF, which we used in the above analysis
(but see Scott et al. 2004). To explore the possible role of the SED on
the deduced NLR parameters we repeated the analysis in Section 3 with
a modified softer SED, where we added a break at 1 Rydberg to a power-law
slope of $-1.5$ extending to 70 Rydberg ($\sim 1$~keV), as suggested
by the far UV to soft X-ray composite of Laor et al. (1997). The rest of
the SED remains unchanged.

The resulting softer SED values, $n_e^s$, and $U^s$, are very
similar to those obtained with the harder MF SED. Apart from two outliers,
where log $U$ changed from $\sim -2.3$ to $\sim -1$, due to the degeneracy
of the $U$ solution at high $U$ (Fig.~3), for the other objects we
find a very tight relation between $U$ and $U^s$. Specifically,
we find a best fit relation
$\log U^s=0.9\log U-0.24$, with an RMS scatter of 0.05, and a Spearman rank order
correlation coefficient $r_S=0.989$. This relation implies an average
systematic offset of
$\log U^s-\log U=-0.04$ to 0.11, for $\log U=-2$ to $-3.5$, which is well within the likely
systematic and statistical errors in our solution for $U$. Similarly, for
$n_e^s$ we find $\log n_e^s=1.02\log n_e-0.17$, an RMS scatter of 0.02,
$r_S=0.996$ (excluding the two outliers), and again a negligibly small offset
of 0.07 to 0.03 for $\log n_e=5$ to 7.
For the soft SED covering factor CF$^s$ we get CF$^s=1.00$CF+0.24, with an RMS
scatter of only 0.01 and $r_S=0.999$. Thus, CF$^s$ is essentially perfectly
correlated with CF, but it is offset to higher values by a factor of
1.74(=$10^{0.24}$), as
expected since the softer SED requires a larger CF to produce a given \Hb\ EW.
Note that the exact absolute value of the best fit parameters, $n_e$, $U$, and CF,
for each object are not crucial for our study of the modulation of the \Oxs\
strength, and what mostly matters is the range of values for these parameters.

\subsubsection{Comparison with Earlier Studies}
Binette et al. (1996, and references therein) discuss evidence for the
presence of significant contribution to the NLR emission from matter
bounded gas. This component produces mostly high ionization lines, and it
is invoked to explain the large relative
strengths of the \He\ and \Civ\ lines, seen in the NLR emission of Seyfert~2
galaxies (e.g. Ferland \& Osterbrock 1986), as well as the strength of other
lines not measured in our
sample. The single and two zone models we use do not include such
a matter bounded gas component. However, we suspect that such a component may not be
required by our data, and the anomalous strength of \He\ may be interpreted
differently, as further discussed below.

Binette et al. (1996) used a mean \Civ/\Hb\ flux ratio of $12\pm 8$ from
Ferland \& Osterbrock (1986). However, using the narrow \Civ\ EW
measurements available
for our sample (BL05), we deduce a significantly lower mean value of $4.7\pm 6$,
which is likely biased to large values as it does not exclude low significance
level narrow \Civ\ measurements. The weakness of the NLR \Civ\ component in
broad line AGN was already noted by Wills et al. (1993) in Hubble Space Telescope
(\hst) spectra of seven radio loud AGN\footnote{The apparently
discrepant \Civ\ NLR emission in Type II/Type I AGN is inconsistent
with the inclination unification schemes for AGN, and this needs to be
addressed in future studies.}. The lower mean \Civ/\Hb\ value is consistent with
standard photoionization model predictions (e.g. Wills et al. 1993;
Groves et al. 2004a), and thus there is no evidence from \Civ\ for
a matter bounded NLR component in our objects.

The mean \He/\Hb\ flux ration in Seyfert 2 galaxies is $0.29\pm 0.1$
(Ferland \& Osterbrock, 1986). We measure a similar mean value of
$0.30\pm 0.12$ in 43 of our objects where a narrow ``\Oxs-like"
\He\ is detectable. As
shown in the extensive set of calculations by Groves et al. (2004a;
2004b), dusty photoionized gas models with $\log U\sim -2$ can
produce \He/\Hb$\sim 0.3$ in ionization bounded gas. The models of Binette et al.
(1996) do not include dust, and these produce lower \He/\Hb\ ratios, which
lead Binette et al. to conclude that an additional population of matter
bounded gas clouds must be present at the NLR.
Dusty photoionized gas with $\log U\sim -2$ produces a higher \He/\Hb\ ratio
since the \Hb\ line originates from deeper layers compared to \He,
and is thus more strongly suppresed by dust absorption. Most of the
line emission in dusty photoionized gas originates within a surface
layer with a column of $\le 10^{21}$~cm$^{-2}$, where the dust opacity
is $\le 1$. The line emission from this surface layer mimics to some
extent the line emission from matter bounded, dust free photoionized
gas, as invoked by Binette et al. (1996).

Dopita et al. (2002) calculated photoionization models for isobaric dusty
gas, and proposed that the ionization parameter in such gas tends to a
constant value, independent of the surface gas
density, once radiation pressure dominates gas presure. Dopita et al.
further suggested this effect explains the small dispersion in the NLR
line ratios (log~\Oxs/\Hb $\sim$0.9-1.2 in their sample), and the implied
nearly constant $\log U\sim -2$ in the NLR. However, here we find a much larger
range (log~\Oxs/\Hb $\sim$0-1.1, Fig.~3), which implies a correspondingly larger
range in $U$. This
argues against the effectiveness of the mechanism proposed by Dopita et al.
to produce a fine tuned $U$ in the NLR.
We note in passing that the dusty isochoric \cl\ model used here produces
nearly identical line ratios to those reported by
Groves et al. (2004a; 2004b) for isobaric dusty gas with significant
radiation pressure. For example,
Groves et al. find log~\Oxf/\Oxs$=-1.65$ for $n_e=10^3$~cm$^{-3}$,
$U=0.1$, and solar abundance (Fig.~13 in Groves et al. 2004b), while
we find $-1.75$ for the same parameters (Fig.~3). At
$U=10^{-4}$, where radiation presure effects on dust are negligible,
their model gives a ratio of $-2.3$, and \cl\ gives an essentially
identical value of $-2.36$.

\subsection{Which parameter modulates most strongly the \Oxs\ EW?}

Figure 5 (upper panels) presents the dependence of the \Oxs\ EW on
$U, n_e$, and CF, for the single zone model.
The \Oxs\ EW ranges over a factor of 40 (4\AA\ to
157\AA) in our sample of 40 AGN with complete detections (compared
with a range of $>300$ in the complete sample of 87 objects). The
\Oxs\ EW is most strongly correlated with the CF. The Spearman
rank order correlation coefficient, $r_S$, for the 40 detected
objects is 0.699, which has a null probability of Pr=$5.2\times
10^{-7}$ (note that the distribution of the \Oxf\ non detections
is consistent with the
distribution of the detections). The CF covers a range of about 10
(1.9 to 20.5 per cent, Table 1), and since the EW of all lines is linear
with CF, the CF modulates the EW by a factor of 10, making it the
dominant factor in modulating the \Oxs\ EW.

The \Oxs\ EW is also significantly correlated with $n_e$
($r_S=-0.473$, Pr~=~$2\times 10^{-3}$), though the \Oxf\
non detections may affect the strength of this correlation if the
true \Oxf\ EW are well below the upper limits. The range of
$n_e\sim 10^5-10^{6.5}$~cm$^{-3}$ amounts
to a modulation of the \Oxs\ EW by an additional factor of 3 (e.g.
the $\log U=-2.5$ curve in Fig.~2, upper panel). Finally, the
\Oxs\ EW shows only a marginal correlation with $U$ ($r_S=-0.384$,
Pr~=~$1.5\times 10^{-2}$, not affected by upper limits).
 Although the range of $U\sim 10^{-3.5}-10^{-2.2}$
modulates the \Oxs\ EW  by a factor of $\sim 4$, its weak correlation
with \Oxs\ EW indicates it mostly contributes to the scatter in the
previous two relations, and does not have a large systematic effect
on the range of \Oxs\ EWs.

The lower panel of Fig.~5 presents the dependence of the \Oxs\
EW/CF on $U, n_e$ and CF. The \Oxs\ EW/CF represents the \Oxs\
emissivity per unit solid angle of the NLR gas. This parameter ranges
over a factor of 5, which represents the amount of \Oxs\ EW
modulation by the combined effects of $n_e$ and $U$. The NLR \Oxs\
emissivity per unit solid angle is strongly correlated with $U$
($r_S=-0.800$, Pr~=~$5.9\times 10^{-10}$) and somewhat less
strongly with $n_e$ ($r_S=-0.708$, Pr~=~$3.2\times 10^{-7}$, again
weakened by non detections). This emissivity is completely
unrelated with CF ($r_S=-0.034$, Pr~=~0.83). We finally note that
$n_e, U$, and CF, are not significantly correlated with each other.

As expected (Section 4.1.3), essentially identical results were
obtained with the modified soft SED. Somewhat stronger correlations
were obtained in the two zone approximation. Specifically, the \Oxs\
EW correlation with $U$ rose to 0.454 and with CF$_{\rm out}$ to 0.778
(both parameters for the outer zone), while
the correlations for the \Oxs\ emissivity (Fig.~5, lower panel) remained
essentially unchanged.

\begin{figure*}
\includegraphics[width=125mm,angle=-90]{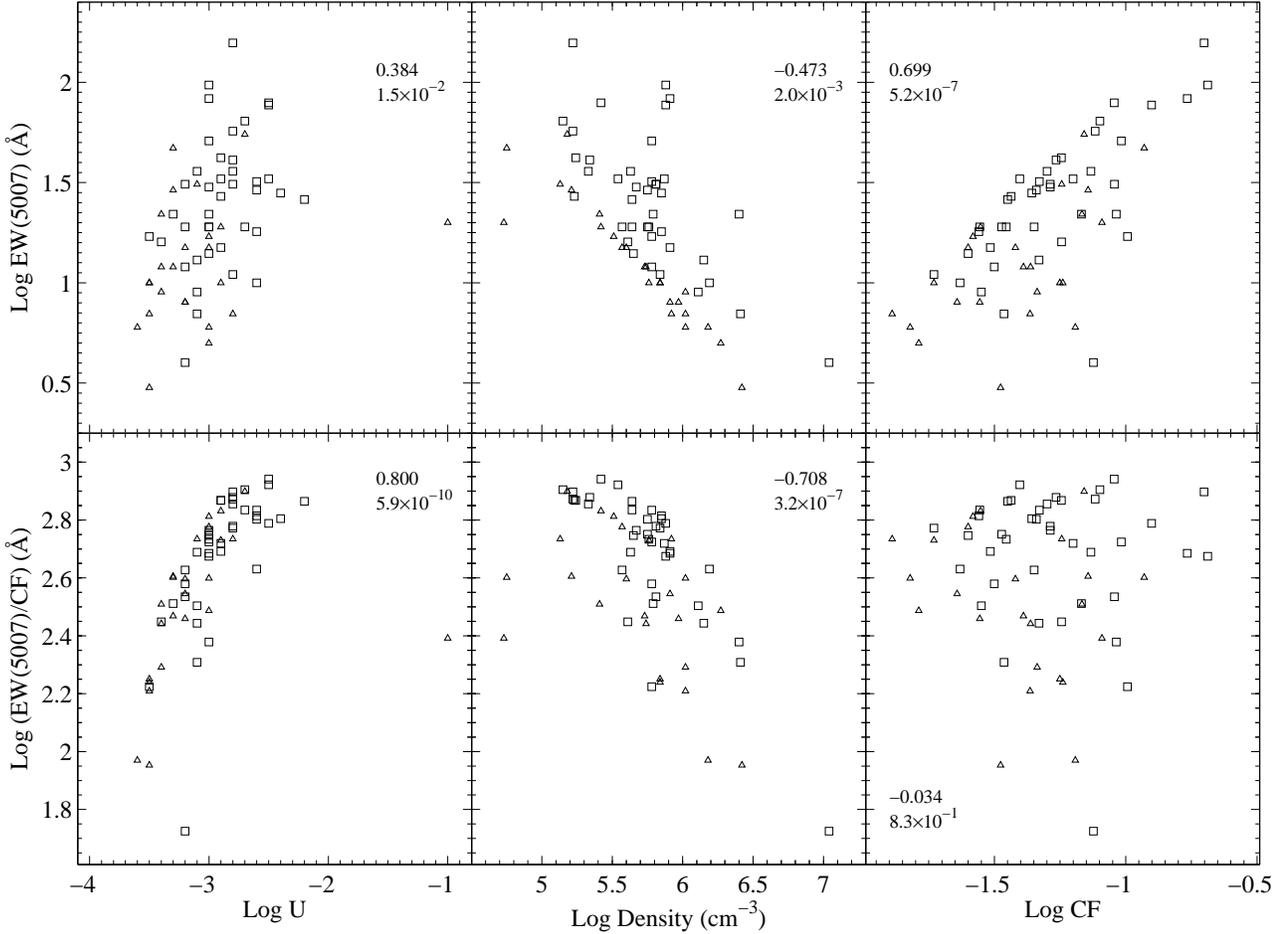}
\caption{The dependence of \Oxs~EW (three upper panels) and
\Oxs~EW/covering factor (three lower panels) on the ionization
parameter, density and covering factor. The Spearman rank-order
correlation coefficient (\rS) and the null probability (Pr) are
indicated in each panel.}
\end{figure*}

\subsection{Relation with EV1 parameters}

The \Oxs\ line is one of the main components of the BG92 EV1 set
of correlations. These correlations may be driven by some
fundamental parameters, such as \Ledd\ (BG92). Below we briefly
explore whether the NLR physical parameters $n_e, U$, and CF are
correlated with some of the non-\Oxs\ EV1 components (as listed in BG92).

The CF is significantly correlated with the absolute $V$ band
magnitude $M_V$ ($r_S=0.512$, Pr~=~$7.6\times 10^{-4}$), such that
the CF decreases with increasing luminosity. This accounts for the
tendency of the \Oxs\ EW to decrease with increasing luminosity in
the BG92 sample (note that Shemmer et al. 2004 do not find this effect
in luminous high-$z$ quasars). The only other parameter
significantly correlated (i.e.
Pr~$<10^{-2}$) with CF is the EW of the broad He~II~$\lambda 4686$
line ($r_S=0.452$, Pr~=~$3.4\times 10^{-3}$). This correlation is
not simple to interpret, and it may be induced by the strong
inverse correlation of the broad He~II~$\lambda 4686$ EW with
luminosity, and the above inverse correlation of luminosity and
CF. Interestingly, there is no significant correlation of CF and
\Ledd\ (taken from BL05; $r_S=-0.133$, Pr~=~0.41). Such a
relation was suggested by BG92 as the physical mechanism driving
the set of EV1 correlations (see also Kraemer et al. 2004).

The $n_e$ is significantly correlated with two surprising
parameters, the soft X-ray slope ($r_S=-0.633$, Pr~=~$1\times
10^{-4}$), and the compact to total radio flux ratio from
Kellermann et al. (1989) ($r_S=0.560$, Pr~=~$1.7\times 10^{-4}$).
A marginally significant correlation exists with the Fe~II/\Hb\
flux ratio ($r_S=0.415$, Pr~=~$7.8\times 10^{-3}$), which may
point at a relation between the NLR density and the BLR metalicity.

The $U$ is significantly correlated with the continuum luminosity
at 3000~\AA, and with the broad \Hb\ line FWHM, leading to a
somewhat stronger, but hard to interpret correlation ($r_S=0.552$,
Pr~=~$2.3\times 10^{-4}$) with the estimated black hole mass
($M_{\rm BH}$, see BL05). A stronger correlation exists with the
Fe~II/\Hb\ flux ratio ($r_S=-0.608$, Pr~=~$3.1\times 10^{-5}$).
Again, this may point at a relation between the NLR ionization level
and the BLR metalicity.

The two zone analysis yields similar correlations for CF$_{\rm out}$
with $M_V$ ($r_S=0.456$, Pr~=~$3.1\times 10^{-3}$), and with the broad
He~II~$\lambda 4686$ EW ($r_S=0.460$, Pr~=~$2.8\times 10^{-3}$).
But, there are no significant correlations with CF$_{\rm in}$. The correlations
for $U_{\rm out}$ are somewhat stronger than those for the single
zone $U$ (with $M_{\rm BH}$, $r_S=0.590$, Pr~=~$6.1\times 10^{-5}$;
with Fe~II/\Hb\  $r_S=0.654$, Pr~=~$4.8\times 10^{-6}$).

\subsection{The size of the NLR}

The distance of the \Oxs\ emitting region from the central ionizing
continuum source, $R_{\rm NLR}$, can be determined from the measured
$n_e$ and $U$ for each object, and the estimated ionizing luminosity
$L_{\rm ion}$, as follows. By definition $U\equiv n_{\gamma}/n_e$,
where the ionizing photon density is given by
$n_{\gamma}=L_{\rm ion}/4\pi R_{\rm NLR}^2c\langle h\nu\rangle$,   where
$\langle h\nu\rangle$ is the mean ionizing photon energy (3.03 Rydberg,
for the \cl\ MF ionizing continuum). These relations give
$R_{\rm NLR}=K L_{\rm ion}^{1/2}(Un_e)^{-1/2}$, where
$K\equiv (4\pi c \langle h\nu\rangle)^{-1/2}$.

Figure 6 presents the luminosity dependence of $R_{\rm NLR}$ for the single
zone and for the two zone approximations. The correlation for the single
zone is strong ($r_S=0.836$, Pr~=~$2\times 10^{-11}$), with
a best fitting relation
\[ R_{\rm NLR}=40L_{44}^{0.45}~{\rm pc}, \]
where $L_{44}=\nu L_{\nu}(4861)/10^{44}$erg~s$^{-1}$.
For the two zone approximation we get a somewhat weaker
but still significant correlation of the size of the outer zone
($r_S=0.648$, Pr~=~$6\times 10^{-6}$),
with a best fitting relation
\[ R_{\rm NLR}^{\rm out}=750L_{44}^{0.34}~{\rm pc}. \]
The assumption of fixed $n_e$ and $U$ values in the inner zone
implies $R_{\rm NLR}^{\rm in}=L_{44}^{0.5}$~pc by construction.

The single zone $R_{\rm NLR}$ versus $L$ relation implies sizes
which are much smaller than deduced through imaging
of spatially extended \Oxs\ emission in Type I AGN with the \hst~
(Bennert et al. 2002; Schmitt et al. 2003). For
example, in four overlapping PG quasars (0026+129; 0053+251;
0953+414; 1307+085) Bennert et al. find
$R_{\rm NLR}=2.2$-5.9~kpc, while we get $R_{\rm NLR}=84$-116~pc,
i.e. about $\sim 40$ times smaller sizes. In the two zone
approximation we get  $R_{\rm NLR}^{\rm out}=$1.3-1.7~kpc, i.e.
within a factor of 2-3 of the Bennert et al. values.
However, the Bennert et al. \hst\ images appear to be measuring the
kpc-scale `extended NLR' (ENLR, Unger et al. 1987),
which according to the surface brightness values in Bennert et al. (Table
2 there), may include a significant component of unresolved
\Oxs\ emission. A more carefull analysis of these images is required
to determine the fraction of unresolved \Oxs\ emission, and the
implied constraints on the \Oxs\ radial emissivity distribution.
Note also that the extrapolation of the Bennert et al. relation to
higher $L$ implies $R_{\rm NLR}$ values which are ruled out by ground
based observations of high $z$ quasars (Netzer et al. 2004), also
pointing at a more compact $R_{\rm NLR}$. Similarly, a compact NLR
($\la 100$~pc), consistent with our estimates, was deduced by
Kraemer et al. (1998), based on detailed photoionization modeling
of the narrow line emission in NGC~5548.

\begin{figure}
\includegraphics[width=84mm]{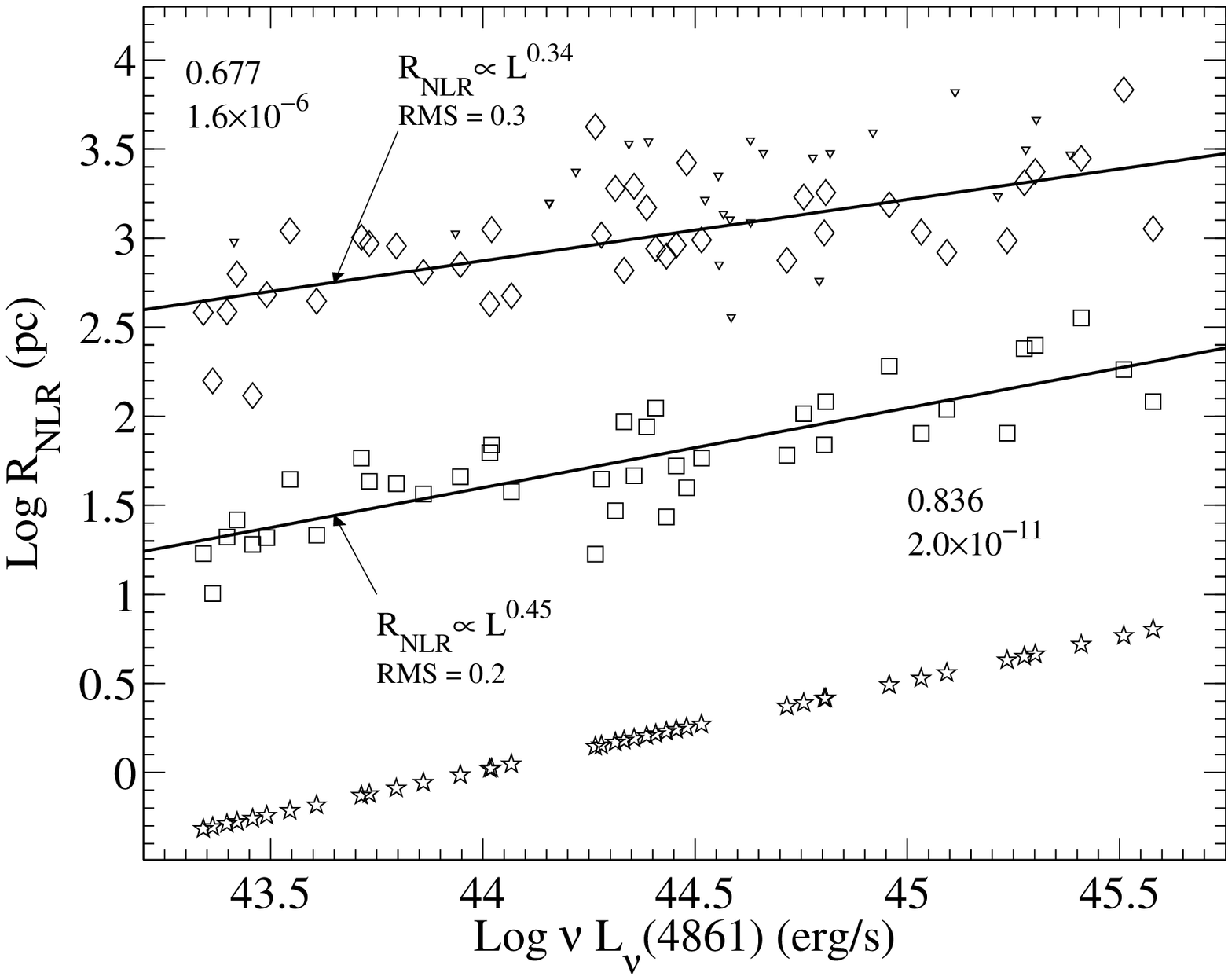}
\caption{The luminosity dependence of $R_{\rm NLR}$
for the single-zone and two-zones models. Note the small
scatter in the relation for the single-zone model (empty squares).
A somewhat larger scatter is present for the outer zone in the
two-zones model (detections-empty diamonds, triangles-upper limits).
The radius of the inner region in
the two-zones models (empty stars in the lower curve)
follows an $L^{1/2}$ relation by construction. The upper curve
indicates that most of the \Oxs\ emission should be resolved
with \hst\ imaging.}
\end{figure}

\begin{figure}
\includegraphics[width=84mm]{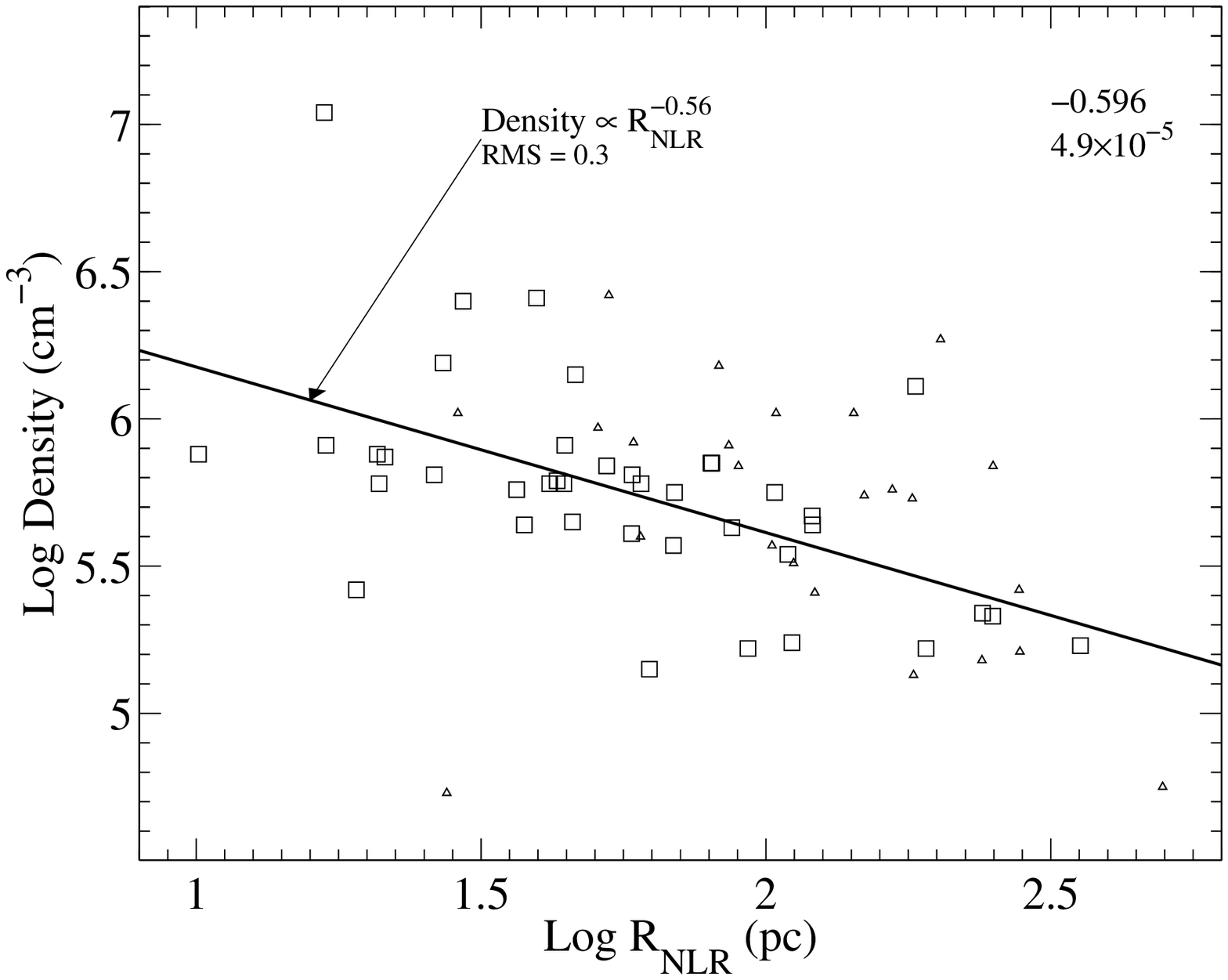}
\caption{The radial gradient in NLR density. The objects with
upper limits may have a larger $R_{\rm NLR}$, and lower density,
which would make this correlation stronger.}
\end{figure}

The slope of 0.45 for the single zone model implies that $Un_e$, i.e.
$n_{\gamma}$ is roughly constant. The scatter in the $R_{\rm NLR}$ versus
$L$ relation is only 0.2 (RMS in log $R_{\rm NLR}$), which is
significantly smaller than expected based on the measured range
of values for $U$ and for $n_e$ \footnote{Phenomenologically, this results from a
weak anti-correlation between $U$ and $n_e$, which produces a
smaller than expected dispersion in $Un_e$, compared to the
dispersion in each parameter.}. The small spread in the $R_{\rm
NLR}$ versus $L$ relation, and a slope close to $0.5$, indicate
a small spread of $n_{\gamma}$ at the \Oxs\ emitting region. It is
not clear why such a small range in $n_{\gamma}$ should be
present~\footnote{The small spread in the BLR $n_{\gamma}$ can be understood
in terms of the critical $n_{\gamma}$ required for dust sublimation
(e.g. LD).}.  The two zone model shows
a somewhat larger scatter (0.3 RMS in log $R_{\rm NLR}^{\rm out}$),
and a flatter slope (0.34), which indicates a non-uniform $n_{\gamma}$.

The tight $R_{\rm NLR}$ versus $L$ relation may be partly induced
by the rejection of objects with non-detected \Oxf. In these objects, the
arbitrary assumption that the true \Oxf\ EW is 0.1 times the upper limit
value leads to
$R_{\rm NLR}$ values which are an order of magnitude larger than
deduced for the rest of the objects. Thus, higher quality optical
spectra, with complete detections of \Oxf, are required in order to
measure more accurately the amount of scatter in the $R_{\rm NLR}$ versus $L$
relation.

Figure 7 shows that there is also a significant inverse
correlation between $R_{\rm NLR}$ and $n_e$ ($r_S=-0.596$,
Pr~=~$4.9\times 10^{-5}$). The slope of the relation is $-0.56$.
This relation is consistent with earlier findings of a negative
radial density gradient in the NLR. In this case, inclusion of
upper limits on \Oxf\ as detections, at a tenth of their value, extends the
correlation to large $R_{\rm NLR}$ and lower $n_e$, indicating
that this correlation cannot be induced by ignoring the non detections.

Inspection of the correlations of $R_{\rm NLR}$ with the EV1
parameters yields that the broad \Hb\ FWHM is also correlated with
$R_{\rm NLR}$ ($r_S=-0.494$, Pr~=~$1.2\times 10^{-3}$). This leads
to a a remarkably strong correlation of $r_{\rm
NLR}$ with $M_{\rm BH}$ ($r_S=0.744$, Pr~=~$3.8\times 10^{-8}$).
This strong correlation stands in contrast with the lack
of correlation of $R_{\rm NLR}$ with \Ledd\ ($r_S=-0.002$,
Pr~=~0.99). We note, however, that $R_{\rm NLR}^{\rm out}$ is not
significantly correlated with $M_{\rm BH}$ ($r_S=0.338$,
Pr~=~$3.3\times 10^{-2}$). We do not have a simple
plausible explanation for the strong $R_{\rm NLR}$ versus $M_{\rm BH}$
correlation. However, as mentioned above, one should note that this correlation
is partly based on the $R_{\rm NLR}$ versus $L$ relation, which needs to be
verified with higher quality spectra allowing complete detections.

\subsection{The CF of the NLR}

Figure 8 presents a comparison of the single zone CF and the two
zone CF$_{\rm in}$ and CF$_{\rm out}$, for both the hard MF SED,
and for the softer $-1.5$ slope SED (Section 4.1.3). In the two
zone model with the soft SED we require the inner zone to produce
a similar \Oxf/\Oxs\ ratio as in the hard SED model, which lead
to $\log U=-1.4$ for the inner zone (rather than $\log U=-1$
with the hard SED). As already noted above (Section 4.1.3), the
softer SED implies CF values which are higher by 1.74
compared to the hard SED. A similar
effect is seen in Fig.~8 for the two zone solution, where
CF$_{\rm in}$ and  CF$_{\rm out}$ in the soft SED solution are
systematically higher than in the hard SED solution.
 For both SEDs
CF and CF$_{\rm out}$ are rather well correlated, but CF$_{\rm in}$
can differ by a factor of 3-5 from CF. This is expected since CF$_{\rm in}$
is largely set by \Oxf, while both CF and CF$_{\rm out}$ are largely
set by \Oxs.

Independent constraints on the radial distribution of the covering
factor of dusty gas in the NLR can be obtained from models for the IR SED of AGN.
In the two zone model, the inner zone is within a factor of few of
the dust sublimation radius (e.g. Fig.~8 in LD), and
the associated dust radiation should thus peak at $\sim 3-5~\mu m$, while
the outer zone is at a $\sim 750$ larger radius, and should thus
peak at $\sim 50-100~\mu m$. The ratio CF$_{\rm out}$/CF$_{\rm in}$
should then determine the shape of the far to near IR SED.
Interestingly, Fig.~8 indicates that this ratio is systematically
different for the soft and the hard SEDs, which suggests one may be
able to obtain constraints on the ionizing SED based on the observed
NLR emission and IR SED. The two zone model presented here is obviously
highly simplified, and a more realistic model should invoke a
continuous radial distribution of dusty gas.

As discussed earlier, the large fraction
of bolometric luminosity emitted in the IR ($\sim 0.3-0.5$ in typical
AGN), requires a correspondingly large CF of dusty gas (e.g. Sanders 1989).
Earlier studies noted that the CF deduced based on the EW of NLR non-resonant
recombination lines is only a few percent, and concluded that there must exist an
additional NLR component with a
large CF and $U\gg 10^{-2}$ (Voit 1992; LD, Netzer \& Laor 1993).
Our single zone model yields
log~CF$\sim -1\pm 0.5$ (for the soft SED), which is still too low.
A somewhat larger total CF is obtained with the two zone model, mainly due
to our assumption of $\log U=-1$ in the inner zone. As briefly mentioned in
Section 2, some of our solutions with $\log U<-2.2$ will be degenerate with
models with $\log U>-1$ (Fig.~3). Such high $U$ models will require a higher
CF, which may be in better agreement with the IR constraints on the CF.
However, the best test for such high $U$ models is through the various
expected high
ionization lines (originating in the surface layer
where dust absorption is negligible).

\begin{figure}
\includegraphics[width=84mm]{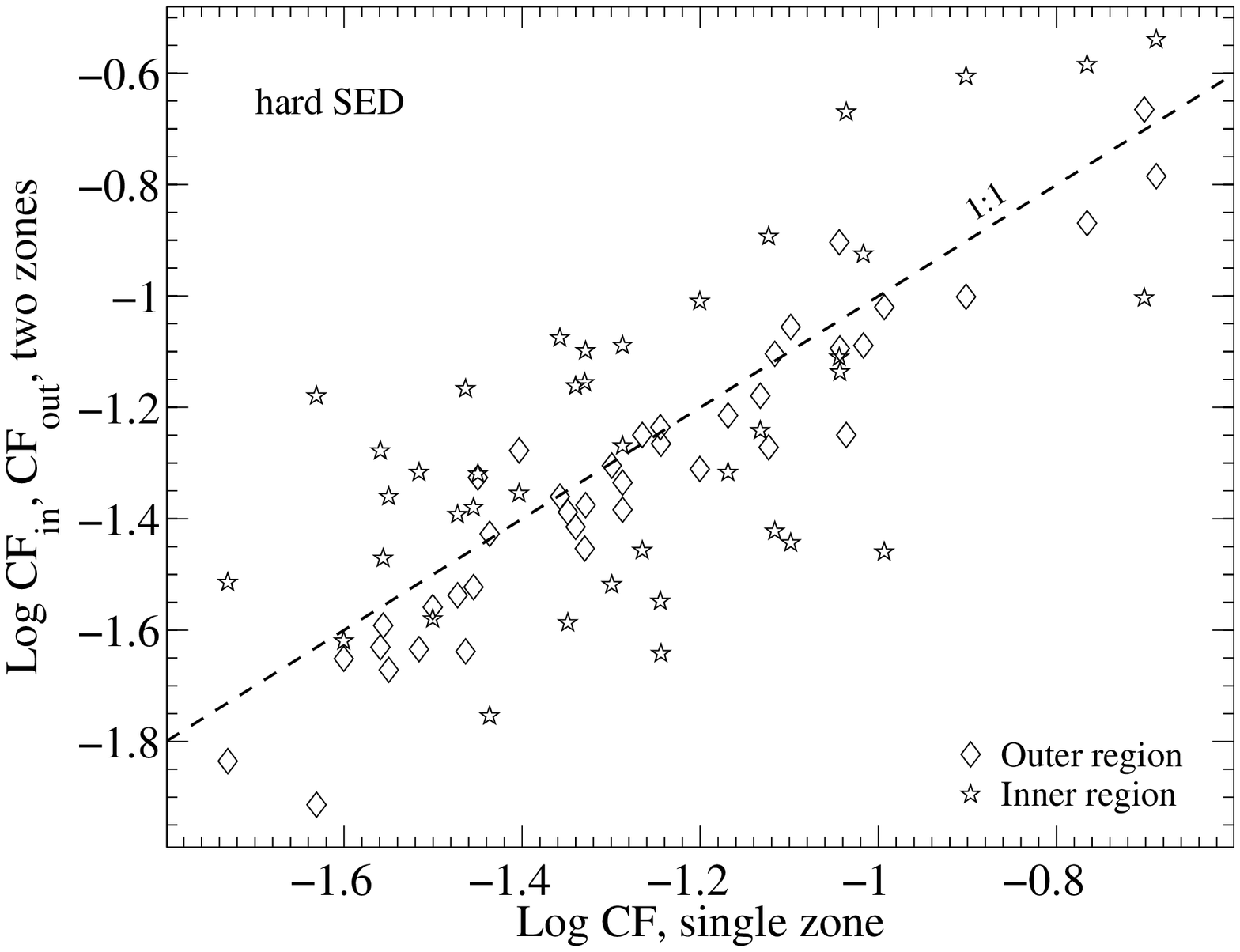}
\includegraphics[width=84mm]{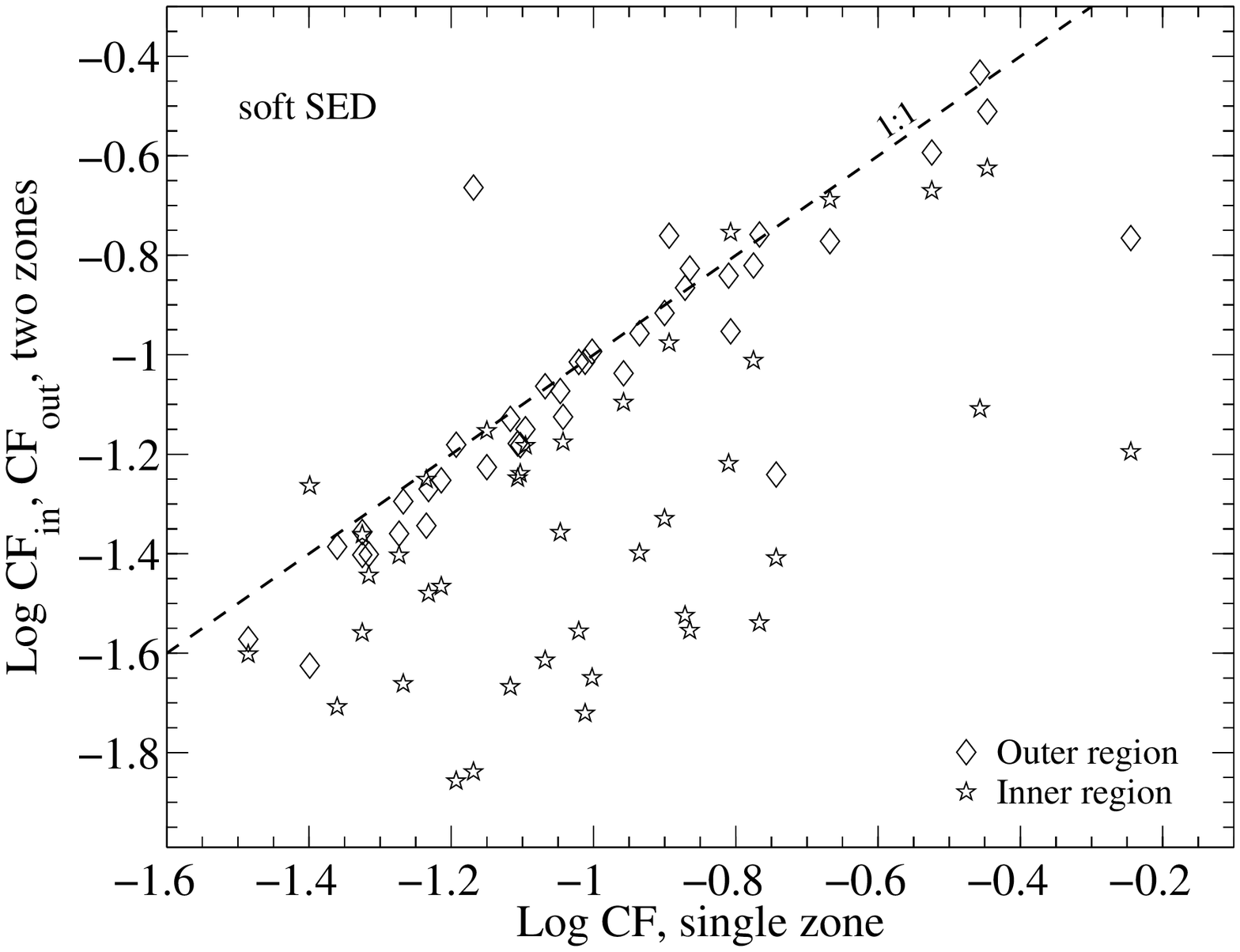}
\caption{The covering factor of the outer (empty diamond) and
inner (empty star) regions in the two-zones model versus the
covering factor of the single-zone model. Upper panel, the
relation for the assumed hard ionizing MF SED. Lower panel, the relation for the
soft ionizing SED. A 1:1 relation is denoted by the dashed line in both panels.
Note that that the ratio CF${\rm in}$/CF${\rm in}$ in a given object,
which sets the observed IR SED, is a function of the ionizing SED.}
\end{figure}

Finally, we note that the \Oxs\ line is one of the main NLR coolants over
a rather wide range of $n_e$ and $U$ (e.g. Ferguson et al. 1997), and thus
should be a good tracer for most of the NLR gas. The \Oxs\ line becomes
a negligible coolant at $\log U<-3.5$ (e.g. Fig.~3), once most O is in
O~I and O~II forms. However, a large covering factor of low $U$ NLR gas
without \Oxs\ emission can be
ruled out in most objects based on the following argument.
Such a low $U$ component would account
for a significant fraction
of the \Hb\ line EW, and thus the inferred \Oxs/\Hb\ ratio
in the \Oxs\ emitting region will
be significantly larger than the observed ratio.
The commonly observed ratio of \Oxs/\Hb$\sim 10$
is already quite close to the maximum possible value in photoionization
models (e.g. Fig.~2, upper panel), and it does not allow more than $\sim 1/2$,
of the \Hb\ line to originate in non \Oxs\ emitting gas.

\section{Conclusions}

We use the \Oxs\ line profile to measure the EW of the \Oxf\ and
\Hb\ emission lines in the BG92 sample of bright $z<0.5$ AGN. All three
lines are
detected in 40 out of the 78 sample objects where optical spectra are available.
The EW of the three lines are used, together with photoionization models,
to infer $n_e$, $U$, and CF of the \Oxs\ emitting gas in the NLR.
We find the main following results:

1. The inferred $n_e$ and $U$ in the single zone approximation
show a relatively small range
($n_e\sim 10^5-10^{6.5}$~cm$^{-3}, U\sim 10^{-2.5}-10^{-3.5}$),
which corresponds to the range where the \Oxs\ emissivity
is maximized.

2. The strength of the \Oxs\ line is mostly modulated by the range
of CF of the photoionized gas. The range of \Oxs\ emissivity per unit
solid angle, set by $n_e$, and $U$, accounts for about half
the range produced by the CF. Similar results are obtained for soft
and hard ionizing SEDs, and for an extreme two zone approximation for the NLR.

3. The NLR CF is inversely correlated with luminosity, consistent
with some earlier suggestions, but it is not correlated with \Ledd, unlike
earlier suggestions.

4. We find a rather tight radius luminosity relation,
$R_{\rm NLR}=40L_{44}^{0.45}$~pc, in the single zone
approximation. We also find an unexpected strong correlation of
$R_{\rm NLR}$ and $M_{\rm BH}$. However, both relations may be
induced by our \Oxf\ detection limits .

5. The observed line ratios can also be fit with a two zone
model, where most of \Oxs\ originates in an $n_e=10^3$~cm$^{-3}$
component at
$R_{\rm NLR}^{\rm out}=750L_{44}^{0.34}$~pc,
and most of \Oxf\ in an $n_e=10^7$~cm$^{-3}$ component at
$R_{\rm NLR}^{\rm in}=L_{44}^{0.5}$~pc. Significant constraints
on the spatial distribution of the \Oxs\ emission can be obtained
through \hst\ imaging of luminous Type I AGN, which should be able to
resolve most of the \Oxs\ emission if it occurs in $n_e\sim 10^3$~cm$^{-3}$
gas.

Significantly higher quality optical spectroscopy of AGN, allowing complete
detections of the \Oxf\ line, is required to establish the strength
of the $R_{\rm NLR}$ correlations with $L$ and $M_{\rm BH}$.

Independent constraints on the radial distribution of the NLR gas
covering factor can be deduced from the IR SED. These constraints,
together with measurements of lower and higher ionization narrow lines,
and more realistic photoionization models involving a
radial distribution of dusty gas (e.g. Ferguson et al. 1997), can provide a
better understanding
of the origin of the observed extreme range in narrow line strengths in AGN.

\section*{acknowledgments}
We thank H. Netzer for the many helpful comments. We also thank
T.~Boroson for providing the optical spectra and accurate
redshifts for all objects, Z.~Shang for providing
optical spectra, M.-P.~V\'{e}ron-Cetty for providing
the I~Zw~1 spectra, and G. Ferland for making the
photoionization code \cl\ publicly available. This research was supported by
THE ISRAEL SCIENCE FOUNDATION (grant \#1030/04), and by a grant from
the Norman and Helen Asher Space Research Institute.

\section*{Note added in proofs}
 Schmitt et al. (2003) surveyed with \hst\ the extended \Ox\ emission
in a sample of 60 nearby Type I/II Seyfert galaxies. They find that
the maximum detectable extent of the \Oxs\ emission
is given by $R_{\rm NLR}=790L_{44}^{0.33}$~pc
(assuming a mean \Oxs~EW of 35~\AA\ to convert their \Oxs\ luminosity
to our continuum luminosity), which is (fortuitously) very close to our
expression for $R_{\rm NLR}^{\rm out}$. They also find that most of the
\Oxs\ emission is significantly more compact, with
half of the flux originating within $R_{\rm NLR}=140L_{44}^{0.21}$~pc,
and typically a factor of two less in Type I AGN. This size
is within a factor of 2-3 of our single zone $R_{\rm NLR}$.
\end{document}